\documentclass[]{interact}

\usepackage{epstopdf}
\usepackage{subfigure}

\usepackage{amsmath,amssymb,amsfonts}
\usepackage{algorithmic}
\usepackage{graphicx}
\usepackage{textcomp}

\usepackage[table]{xcolor}

\usepackage{natbib}
\bibpunct[, ]{(}{)}{;}{a}{}{,}
\renewcommand\bibfont{\fontsize{10}{12}\selectfont}

\theoremstyle{plain}

\theoremstyle{definition}

\theoremstyle{remark}

\begin{document}


\title{A Clustering Approach for Remotely Sensed Data in the Western United States}

\author{
\name{Ghazal Farhani  \thanks{CONTACT G.Farhani. Email: ghazal.farhani@nrc-cnrc.gc.ca} }
\affil{National Research Council Canada, London, Ontario}}


\maketitle

\begin{abstract}
The increasing frequency and scale of wildfires carry significant ecological, socioeconomic, and environmental implications, prompting the need for a deeper grasp of wildfire characteristics. Essential meteorological factors like temperature, humidity, and precipitation wield a crucial impact on fire behavior and the estimation of burned areas. This study aims to unravel the interconnections between meteorological conditions and fire attributes within the Salmon-Challis National Forest located in east-central Idaho, USA. Through the utilization of remotely sensed data from the Fire Monitoring, Mapping, and Modeling system (Fire M3) alongside meteorological variables recorded between 2010 and 2020, an exploration is conducted into varied meteorological patterns associated with wildfire events. By integrating the computed burned area into the clustering process, valuable insights are gained into the specific influences of fire weather conditions on the extent of burned areas. The Salmon-Challis National Forest, encompassing more than 4.3 million acres and encompassing the largest wilderness area in the Continental United States, emerges as a pivotal research site for wildfire investigations. This work elucidates the data attributes employed for clustering and visualization, along with the algorithms employed. Additionally, the study presents research findings and delineates potential future applications, ultimately contributing to the advancement of fire management and mitigation strategies in regions prone to wildfires.

\end{abstract}


\section{Introduction}

The annual extent of burned areas induced by wildfires ranges from 360 to 380 million hectares (Mha) \citep{chuvieco2016new}. The escalation in wildfire frequency across multiple regions is attributed to anthropogenic climate change \citep{flannigan1988study, flannigan2013global}. Heightened forest fire activity in recent decades has been experienced in the Western United States, resulting in extensive forest mortality, carbon emissions, and degraded air quality \citep{kasischke2006recent, holden2018decreasing}. Additionally, an annual increase in the occurrence of large wildfires and the total burned area has consistently been demonstrated since 1984, with an expansion of 355 square kilometers \citep{dennison2014large}. In tandem, the prevalence of ``Mega-fires'', has notably grown in the western U.S., particularly in Idaho \citep{freyberg2022idaho}.

Adverse effects on soil productivity, an increased potential for invasive plant encroachment, and soil erosion are encompassed by the ecological impact of high burn severity \citep{robichaud2000evaluating, certini2005effects}. Furthermore, significant impacts on land resources, socioeconomic systems, and ecological equilibrium have been imposed by the amplified size and frequency of wildfires \citep{shvidenko2011impact, mcwethy2019rethinking}. Consequently, numerous studies have been undertaken by researchers to improve wildfire detection, prediction, and understanding of fire behavior \citep{finney1994farsite, green1995fire, kreye2014fire}.

A pivotal role in fire behavior and burned area estimation and prediction is played by meteorological variables. Empirical links have been established between fire activities and fire weather variables, such as temperature and relative humidity. Contributing factors to increased area burned in the western U.S. include reduced winter snowpack, increased summer temperatures, and declining summer precipitation and wetting rain days \citep{holden2018decreasing}. Strong correlations have been highlighted by studies exploring the influence of meteorological variables on fire activity between extended periods without rain and low relative humidity with increased burned area \citep{flannigan1988study}. Additionally, air temperature and precipitation, have been associated with long-term fire trends and are known as the primary drivers of total burned area \citep{koutsias2012relationships, giannaros2021climatology}.

The Canadian Fire Weather Index (FWI) System is recognized as one of the most commonly used indices to capture the influence of weather and climate on fire danger and as a predictor of burned area \citep{van1987development}. For example, up to 80\% of the variance in the area burned and number of fires in Portugal has been demonstrated to be explained by the FWI index \citep{carvalho2008fire}. Furthermore, FWI index, employed individually in a recent study, reaffirmed that warmer and drier climates promote extreme weather conditions \citep{giannaros2021climatology}.

While the relationship between weather conditions and forest fires is well-established, the intercorrelation between weather variables and their impact on burned areas remains relatively unexplored. Of great interest is understanding the patterns of meteorological conditions associated with different fire behaviors, shedding light on conditions conducive to rapid fire spread, extreme fire intensity, and the extent of burned area. Hence, in the present study, visualization and clustering methods are employed to enhance the understanding of meteorological conditions and their relationship to the expected burned area. The clustering process is enriched by integrating the estimated burned area, providing novel insights into the burned area under specific fire weather conditions.

The focus of the study is the Salmon-Challis National Forest, situated in east-central Idaho within the Western United States. Encompassing over 4.3 million acres and boasting the largest wilderness area in the Continental United States, this vast forest has seen several fires \citep{sncf}. Leveraging data from the Fire Monitoring, Mapping, and Modeling system (Fire M3), wildfires occurring between 2010 and 2020 in the Salmon-Challis National Forest are investigated to ascertain the role of meteorological variables in shaping the patterns of burned areas.

The research centers on the identification of data clusters based on temperature, relative humidity, wind speed and direction, precipitation, the FWI, and burned area. The application of the t-distributed Stochastic Neighbor Embedding (tSNE) algorithm for dimension reduction facilitates the visualization of patterns in the data, while the Density-Based Spatial Clustering of Applications with Noise (DBSCAN) algorithm assists in the identification of separate clusters. The clusters identified offer intriguing insights into how different extents of burned areas can result from various combinations of meteorological variables.

The potential to enhance fire management, benefiting both natural environments and human communities, particularly in regions prone to wildfires like the Salmon-Challis National Forest, is held by the insights gleaned from the study. An overview of the paper's structure is provided as follows:
In Section 2, the data used and its characteristics are briefly described. The methods employed for clustering and visualizing the data are also explained. Section 3 presents the results of the data clustering method and shares the estimates for the FWI. A comprehensive discussion of the findings is delved into in Section 4. A roadmap for the application of the grouping methods to other areas in the future is also provided.

\section{Data and Methods}

\subsection{Study Area: The Salmon-Challis National Forest}

Encompassing 4.3 million acres across five separate units in east-central Idaho, the Salmon-Challis National Forest encompasses the expansive Frank Church-River of No Return Wilderness Area. This contiguous wilderness area, spanning 2.3 million acres, holds the distinction of being the largest of its kind in the Continental United States \citep{sncf}. Wildfires have been a natural part of the forest's history, with some areas experiencing up to five fires in the past 120 years. However, recent years have seen a surge in the frequency of fires, affecting regions that were rarely impacted by fire in the past. These fires vary in origin, with some initiated by human activities during favorable summer conditions, leading to substantial size, such as the Moose fire in 2022, while others are of natural origin, sparked by lightning strikes, like the Owl fire and the Woodtick Fire in 2022 \citep{sncf}.

The mountainous terrain of the region adds to the challenge and danger of firefighting, as many active fires occur in steep and rugged areas \citep{sncf}. Understanding weather factors becomes crucial in estimating the extent of the burned area and predicting fire size. This knowledge can aid in better firefighting strategies and control efforts for mitigating the impact of wildfires.

\subsection{The fire monitoring, mapping, and modeling system (Fire M3)}

The primary objective of the Fire M3 is to use satellite remote-sensing data to monitor daily fire activity to estimate daily and annual area burned, model fire behavior and biomass consumption, and produce fire maps to model the impact of fire. A hotspot is a satellite image pixel with high intensity in the infrared spectrum that shows heat sources. Hotspots from known industrial sources are removed so that the remaining hotspots show vegetation fires. A hotspot can be for one fire or be one of several hotspots representing a larger fire. The fire M3 hotspots are obtained from multiple sources including the Advanced Very High-Resolution Radiometer (AVHRR) imagery, the Moderate Resolution Imaging Spectroradiometer (MODIS) imagery, and the Visible Infrared Imaging Radiometer Suite (VIIRS) imagery. More details can be found in \citep{cwfis}. 

The Fire M3 model is capable of identifying pixels containing fire (hotspots) and smoke plums from fires. By combining annual hotspot maps with the observed annual changes in a vegetation index from the Systeme Probatoire d'Observation de la Terre (SPOT) Vegetation (VGT) sensors the estimation of the area burned is possible. The annual comparison of the vegetation index for each pixel is done and the significant drop in the index that is coupled to a hotspot is mapped as a burned area. The approach is employed to provide a coarse resolution (based on a 1-km resolution satellite imagery) burned area at the end of each year. As more than 95\% of forest fires exceed more than 10$\,km^2$ the described coarse resolution approach is sufficient for mapping the majority of burned areas \citep{fraser2000hotspot}.


\subsection{Attributes for Fire M3 Hotspots}
The complete list of fire M3 hotspots can be located in \cite{cwfis}, and those attributes that were utilized in the analysis of the current study include temperature (T) in degrees, relative humidity (rh) as a percentage, wind direction (wd) in degrees, wind speed (ws) in $\,km/h$, precipitation (pcp) in mm, the fire weather index (FWI) components, and burned area in hectares. T, rh, wd, ws, and pcp are measured during local noon at hotspot locations. The components of FWI are elaborated upon in \ref{FWI}.

\subsection{FWI} \label{FWI}

The Canadian FWI System \citep{van1987development} is a subsystem of the Canadian Fire Danger Rating System \citep{stocks1989canadian} that includes five components: fine fuel moisture code (FFMC), duff moisture
code (DMC), drought code (DC), initial spread index (ISI), and build-up index (BUI). Here, we briefly, explain each of the components. A complete description can be found in \citep{van1987development, wotton2009interpreting}. 
\begin{itemize}
    \item{FFMC}:
At the beginning of a forest fire, it is strongly influenced by the moisture of consumed fuels on the surface of the forest floor. The moisture content of surface fuels under the shade of a forest canopy is described by the FFMC. The value of the FFMC increases with increasing dryness and it varies from 0 (250\% moisture content) to a maximum of 101 (that is when the litter layer is completely dry). This scale was adopted so that higher numbers of the code represent greater fighter danger.  
    \item{DMC}:
describes the amount of moisture in the upper layers of the forest floor where the organic material is decomposed. The DMC values start from 0 and have no higher bound however, rarely values above 150 have been reported.  
    \item{DC}:
is an index showing the influence of long-term drying on the fuels in the forests. As with FFMC and DMC the increased value of DC means the increased level of dryness of the deep forest floor. In the FWI system, a DC value of 0 shows a saturation moisture content of 400\% and has no maximum value, however, typically it rarely exceeds values over 1000.
    \item {BUI}: It is the combination of the current DMC and DC to estimate the potential heat release in heavier fuels. It is unitless and open-ended.
    \item {ISI: defines based on the integration of FFMC and the observed wind speed and is a unitless value which indicates the potential rate of spread of a fire.}
\end{itemize}

In particular, the threshold value of 30 for FWI is a good indicator of the severity of the fire, it has been shown that values exceeding this value can be strong indicators of the occurrence of extreme fire dangers \citep{bedia2014forest}.

\subsection{Sensitivity of FWI}

In data-driven approaches, it is common to exclude attributes derived from other attributes. Hence, in our analysis, we only utilized the FWI without including its components, namely FFMC, DMC, DC, BUI, and ISI. Of note, in the previous studies, this exclusion was not considered \citep{carvalho2008fire}. However, we were curious about the sensitivity of FWI to each component, so we employed a Random Forest (RF) algorithm. RF is a non-parametric machine learning algorithm that is a suitable choice for finding nonlinear and complex relations between variables and the targets \citep{breiman2001random}. The RF model used FFMC, DMC, DC, BUI, and ISI as predictor variables and FWI as the target variable. The training data encompassed the period from 2010 to 2019, and the model's performance was tested on data from 2020 to assess its goodness. This process aimed to investigate whether FWI is entirely explainable by the five mentioned components, hence allowing us to exclusively use FWI in the clustering process. Additionally, the analysis provided insights into which component plays a major role in determining the FWI index.

\subsection{Clustering Algorithms}
A clustering algorithm is a data-driven approach in unsupervised learning, where data points are grouped together based on similarity without specific labels \citep{unsupervised}. In recent years clustering and visualization methods have gained popularity in remote sensing research \citep{halladin2019t, song2019improved, farhani2021classification}. While there exist several methods, for visualization and pattern discovery within data, this study utilizes the t-distributed Stochastic Neighbor Embedding (t-SNE) method, a popular nonlinear dimensionality reduction technique \citep{maaten2008visualizing}. Additionally, the Density-based spatial clustering of applications with noise (DBSCAN) algorithm is employed, which clusters data points based on density. Here, we provide a brief introduction to both t-SNE and DBSCAN algorithms.

\subsection{t-distributed Stochastic Neighbour Embedding (t-SNE)}
In t-SNE, the objective is to map data points into a low-dimensional space while preserving their pairwise similarities. To achieve this, the probability that data point $i$ selects data point $j$ as its neighbor is calculated as follows \citep{maaten2008visualizing}:

\begin{equation}
P_{i,j} = \frac{exp(-d_{i,j}^2)}{\sum_{k \neq i} exp(-d_{i,k}^2)}
\end{equation}

Here, $P_{i,j}$ represents the probability of data point $i$ selecting data point $j$ as its neighbor. The squared Euclidean distance between two points in the high-dimensional space is denoted as $d_{i,j}^2$ and is defined as:

\begin{equation}
d_{i,j}^2 = \frac{|x_{i} - x_{j}|^2}{2\sigma_{i}^2}
\end{equation}

The parameter $\sigma_{i}$ is determined such that the entropy of the distribution becomes $\log \kappa$, where $\kappa$ is referred to as the ``perplexity''. This perplexity is set by the user as a hyperparameter.

The ultimate goal of t-SNE is to map each data point $x_{i}$ from the high-dimensional space to a lower-dimensional point $y_{i}$. To find the best match between the original distribution's $P_{i,j}$ and its counterpart $q_{i,j}$ in the lower dimension, where $q_{i,j}$ is defined as:

\begin{equation}
q_{i,j} = \frac{exp(-|y_{i}-y_{j}|^2)}{\sum_{k \neq i} exp(-|y_{i}-y_{k}|^2)}
\end{equation}

The t-SNE algorithm uses the Kullback-Leibler divergence as the cost function to minimize the dissimilarities between $P_{i,j}$ and $q_{i,j}$:

\begin{equation}
\text{cost} = \sum_{i} \text{KL}(P_{i}||Q_{i}) = \sum_{i}\sum_{j} p_{j|i} \log \frac{p_{j|i}}{q_{j|i}}
\end{equation}

Here, $P_{i}$ is the conditional probability distribution of all data points given data point $x_{i}$, and $Q_{i}$ is the conditional probability distribution of all data points given data point $y_{i}$. Detailed on the algorithm can be found in \citep{maaten2008visualizing}.

\subsection{Density-based spatial clustering of applications with noise (DBSCAN)}
DBSCAN is a density-based clustering method that is well-suited for identifying clusters within datasets of arbitrary shapes. Unlike other clustering methods like K-means, there is no need to predefine the number of clusters. The input parameters, minPts and $\epsilon$ (epsilon), play crucial roles in the process.

For a point $p$ in the dataset $D$, $\epsilon$ is defined as:

\begin{equation}
N_\epsilon(p) = {q \in D \mid \text{dist}(p, q) \leq \epsilon}
\end{equation}

Here, $N_{\epsilon(p)}$ represents the set of points within the distance $\epsilon$ from $p$, where $p$ and $q$ are two points in the dataset $D$, and dist($p, q$) is the distance function used. Once minPts and $\epsilon$ are set, the clustering process begins.

During the clustering process, data points are categorized into three groups: core points, (density) reachable points, and outliers:

Core point: A point $A$ is a core point if there are at least minPts points (including $A$) within a distance of $\epsilon$.

Reachable point: A point $B$ is considered reachable from a core point $A$ if there exists a path ($P_1, P_2, ..., P_n$) from $A$ to $B$ (with $P_1 = A$). All points along the path, except possibly $B$, are core points.

Outlier point: A point $C$ is classified as an outlier if it is not reachable from any other point.

In this approach, the clustering process proceeds as follows: At each step, an arbitrary point is selected, and following the steps mentioned above, the neighboring points are retrieved. If the resulting group of points forms a cluster with enough points (determined by minPts), it is accepted as a cluster. A complete description of the algorithm is presented in \cite{ester1996density}. 

\section{Result}
\begin{figure}
    \centering
    \includegraphics[scale=.50]{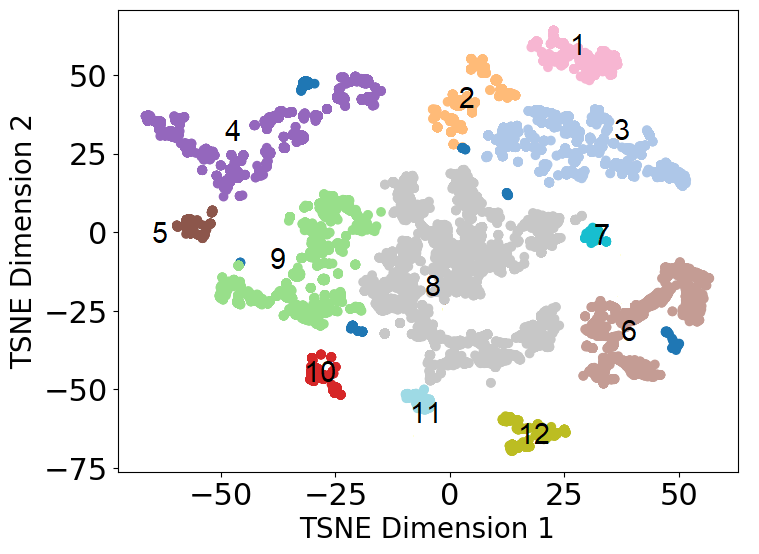}
    \caption{From the tSNE representation of data, 12 distinct clusters are generated.}
    \label{fig:tSNE}
\end{figure}

Between 2010 and 2020, the Salmon-Challis National Forest underwent a process where a t-SNE algorithm was utilized to reduce the data's dimensions to two, enabling easier visualization (refer to Fig~\ref{fig:tSNE}). Following this, the application of DBSCAN led to the identification of 12 distinct clusters. In Table 1, the computed average values of features employed by t-SNE are presented. While FFMC, DMC, DC, ISI, and BUI were not directly used within the t-SNE algorithm, the mean values corresponding to each cluster are provided.

Cluster eight harbors the largest populace, comprising approximately 8000 data points. The average estimated area within this cluster spans 20 hectares, accompanied by an FWI of 36, signifying extreme fire weather conditions. As indicated in the Table, the typical burned area varies between 20 to 60 hectares.

An intriguing pattern emerges in the cluster analysis, notably in clusters five and ten. These clusters exhibit relatively low average temperatures of 18 degrees Celsius, coupled with comparably lower FWI values (contrasted with other clusters). Nevertheless, their burned areas are akin to those with higher temperatures and FWI values.

Another noteworthy observation pertains to clusters two and three. Despite having akin average burned area values, these clusters diverge in their mean temperature, relative humidity, precipitation, and notably, FWI values. This underscores the significance of wind speed and direction in FWI calculations, serving as the primary distinctions between these clusters.

Clusters six and seven similarly share estimated area ranges, yet they markedly differ in all aspects except temperature.

\begin{table}[]
\scalebox{0.8}{
\begin{tabular}{|
>{\columncolor[HTML]{FFFFFF}}l |
>{\columncolor[HTML]{FFFFFF}}l |
>{\columncolor[HTML]{FFFFFF}}l |
>{\columncolor[HTML]{FFFFFF}}l |
>{\columncolor[HTML]{FFFFFF}}l |
>{\columncolor[HTML]{FFFFFF}}l |
>{\columncolor[HTML]{FFFFFF}}l |
>{\columncolor[HTML]{FFFFFF}}l |
>{\columncolor[HTML]{FFFFFF}}l |
>{\columncolor[HTML]{FFFFFF}}l |
>{\columncolor[HTML]{FFFFFF}}l |
>{\columncolor[HTML]{FFFFFF}}l |
>{\columncolor[HTML]{FFFFFF}}l |
>{\columncolor[HTML]{FFFFFF}}l |l}
\cline{1-14}
\cellcolor[HTML]{ECF4FF}Cluster &
  \cellcolor[HTML]{ECF4FF}Population &
  \cellcolor[HTML]{ECF4FF}T &
  \cellcolor[HTML]{ECF4FF}RH &
  \cellcolor[HTML]{ECF4FF}WS &
  \cellcolor[HTML]{ECF4FF}WD &
  \cellcolor[HTML]{ECF4FF}PCP &
  \cellcolor[HTML]{ECF4FF}FFMC &
  \cellcolor[HTML]{ECF4FF}DMC &
  \cellcolor[HTML]{ECF4FF}DC &
  \cellcolor[HTML]{ECF4FF}ISI &
  \cellcolor[HTML]{ECF4FF}BUI &
  \cellcolor[HTML]{ECF4FF}FWI &
  \cellcolor[HTML]{ECF4FF}Area &
   \\ \cline{1-14}
1  & 1274 & 30 & 20 & 9    & 227  & 0.15   & 95 & 186 & 913 & 15.2 & 246 & 50 & 60 &  \\ \cline{1-14}
2  & 1274 & 27 & 24 & 6    & 265  & 0.02   & 94 & 192 & 666 & 10.2 & 220 & 40 & 41 &  \\ \cline{1-14}
3  & 3445 & 26 & 17 & 14   & 218  & 0.02   & 96 & 195 & 912 & 20.7 & 245 & 60 & 42 &  \\ \cline{1-14}
4  & 3461 & 27 & 27 & 0.03 & 0.11 & 0.04   & 93 & 200 & 673 & 7.5  & 223 & 32 & 45 &  \\ \cline{1-14}
5  & 490  & 18 & 55 & 0.25 & 0.15 & 0.30   & 86 & 222 & 710 & 3.6  & 245 & 20 & 38 &  \\ \cline{1-14}
6  & 3187 & 23 & 26 & 6    & 0.50 & 0.01   & 94 & 123 & 571 & 10.5 & 160 & 38 & 28 &  \\ \cline{1-14}
7  & 228  & 24 & 20 & 10   & 81   & 0.0001 & 95 & 222 & 675 & 14.8 & 245 & 50 & 27 &  \\ \cline{1-14}
8  & 7920 & 20 & 28 & 10   & 190  & 0.95   & 93 & 115 & 508 & 11.5 & 141 & 36 & 20 &  \\ \cline{1-14}
9  & 3654 & 23 & 30 & 7    & 233  & 0.08   & 93 & 128 & 564 & 9.7  & 160 & 36 & 28 &  \\ \cline{1-14}
10 & 658  & 18 & 54 & 6    & 337  & 0.10   & 88 & 237 & 734 & 4.7  & 260 & 23 & 42 &  \\ \cline{1-14}
11 & 442  & 21 & 37 & 7    & 0.00 & 0.03   & 92 & 140 & 630 & 8.8  & 178 & 35 & 29 &  \\ \cline{1-14}
12 & 228  & 24 & 32 & 3    & 1.1  & 0.007  & 92 & 130 & 623 & 7.2  & 169 & 30 & 27 &  \\ \cline{1-14}
\end{tabular}} \label{tSNE-table}
\caption{The average values of tSNE-utilized features are displayed. }
\end{table}

\begin{figure}
    \centering
    \includegraphics[scale=.40]{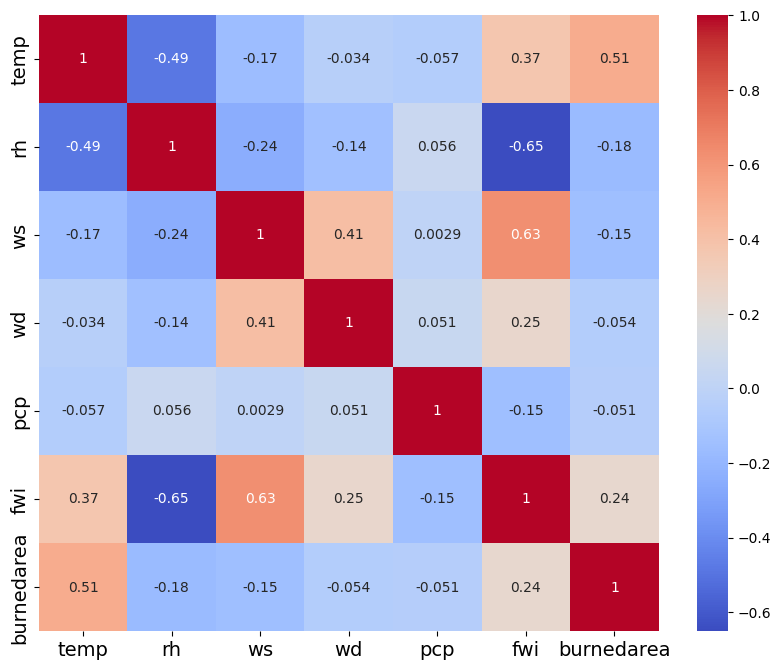}
    \caption{The heatmap visually presents the correlation coefficients among the features utilized in the clustering process.}
    \label{fig:heatmap}
\end{figure}

Given the discernible insights from the clusters, it becomes apparent that distinct combinations of features can lead to comparable burned areas. This prompts an exploration into the scrutiny of intercorrelations among these features, utilizing the Pearson correlation technique. The outcomes, depicted in Fig~\ref{fig:heatmap}, unveil significant positive correlations between burned area and temperature, as well as with the Fire Weather Index (FWI). Conversely, a negative correlation is observed between burned area and wind speed, as well as between burned area and relative humidity. Similarly, FWI demonstrates a marked positive correlation with wind speed and temperature, while concurrently exhibiting a negative correlation with relative humidity. 

\subsection{FWI prediction}
Within the clustering methodology, the FWI metric was harnessed, a composite derived from the product of FFMC, DMC, DC, ISI, and BUI. This led to the exploration of whether it is feasible to predict FWI values without the use of pre-defined functions, relying solely on historical records encompassing FFMC, DMC, DC, ISI, and BUI. This endeavor gains heightened significance due to FWI's pivotal role in the clustering process. Central to the investigation was an exploration into FWI's responsiveness to each of its constituent components, aiming to discern the most influential contributors in an equation-free FWI estimation.

\begin{figure}
    \includegraphics[scale=.33]{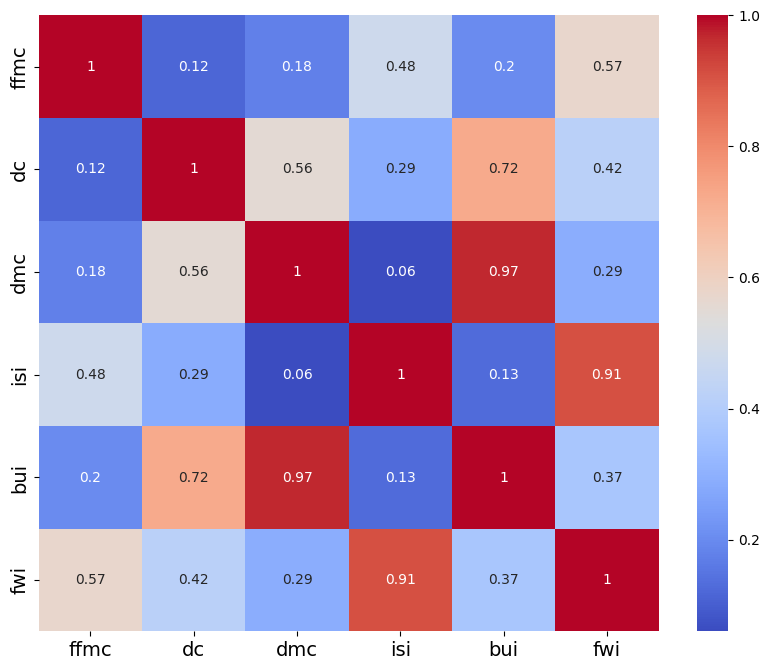}
    \includegraphics[scale=.40]{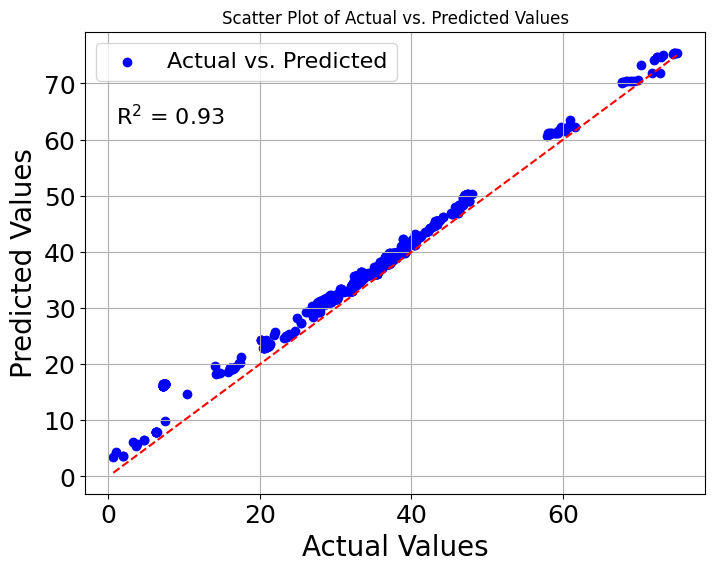}
    \caption{On the left panel, the heatmap visually presents the correlation coefficients among the features utilized in the clustering process. On the right panel, a scatter plot depicting the actual (ground truth) FWI values against the predicted values are shown. The red dashed line indicates the optimal fit, capturing the relationship with a high R$^{2}$ value of 0.93}
    \label{fig:heatmap_FWI}
\end{figure}

The approach involved utilizing a training dataset spanning the years 2010 to 2019. A robust RF regression model was constructed specifically for the prediction of FWI. Prior to commencing model training, the Pearson correlation method was employed to uncover interdependencies among the features. Notably, the observations led to the omission of DMC from the feature set due to its substantial correlation with BUI. This pre-training analysis is visualized in Fig~\ref{fig:heatmap_FWI}, left panel. Moreover, the scrutiny of the heatmap provided unequivocal evidence of FWI's profound correlation with ISI.

Following this preparatory phase, the RF model underwent training. This meticulously crafted model underwent a stringent test on previously unseen data from the year 2020. The outcomes were remarkably compelling, with the predicted FWI values exhibiting a high degree of concordance with the actual ground truth. This assertion is supported by a notable $R^2$ score of 0.93, highlighting the model's efficacy in accurately predicting FWI values. Refer to Fig~\ref{fig:heatmap_FWI}, right panel, for a graphical representation of these outcomes. The core objective behind FWI modeling was to investigate the role of individual components of FWI. Through the analysis of feature importance, it was found that: a substantial portion, exceeding 85\%, of FWI's variability is ascribed to ISI, with an additional 12\% stemming from BUI. Consequently, ISI assumes a dominant position in shaping FWI, corroborating observations in the Pearson correlation analysis (see Fig~\ref{fig:heatmap_FWI}). Furthermore, Table 1 underscores this correlation by distinctly presenting the positive association between FWI and ISI.

\section{Summary and Conclusion}
This study explores a decade-long dataset spanning from 2010 to 2020, derived from fire M3 hotspots. The objective is to construct an extensive wildfire clustering analysis tailored to the Salmon-Challis National Forest context. The focus of this analysis involves utilizing the tSNE and DBSCAN algorithms to cluster wildfire-related data, encompassing parameters such as temperature, relative humidity, wind speed and direction, precipitation, Fire Weather Index (FWI), and the extent of burned area.

Guided by a prior investigation conducted in Portugal \citep{carvalho2008fire}, which highlighted the significance of FWI, BUI, DC, DMC, and FFMC in assessing burned area, this study delves deeper into this relationship. However, rather than using these drivers of FWI directly as inputs for the clustering algorithm, they are employed as predictors for FWI. Notably, the feature importance analysis reveals the prominence of ISI and BUI as key contributors to FWI, while DC and FFMC exhibit minimal influence. Remarkably, ISI captures a significant 84\% of the variance associated with FWI, underscoring its critical role in predicting FWI. This finding sheds light on the pivotal importance of early fire propagation dynamics in shaping the overall fire weather index, offering valuable insights for effective fire management strategies.

Furthermore, consistent with the prior study conducted by \textit{Holden et al.} \cite{holden2018decreasing} in the Western United States, the cluster analysis highlights the correlation between high temperature, low relative humidity, and Fire Weather Index (FWI) with the extent of burned area. In contrast to earlier research \citep{koutsias2012relationships, giannaros2021climatology}, the analysis indicates that precipitation itself does not emerge as a significant driver of total burned area. Instead, the role of wind speed in influencing the burned area becomes evident. The impact of wind speed extends to FWI prediction, where ISI emerges as the most crucial component, itself derived from wind speed.

While the role of high temperature in estimating burned area remains pivotal, a detailed cluster analysis unveils a fascinating insight: even under mild temperature conditions with reduced FWI values, the burned area's extent remains comparable to other clusters. This underscores the intricate interplay of additional parameters in shaping clusters and impacting burned area, underscoring the complex relation between different meteorological variables. Recognizing the considerable influence of parameters beyond temperature and FWI on burned area emphasizes the necessity for adaptable and flexible suppression strategies.

In summary, the clustering analysis reveals diverse potential fire patterns within the Salmon-Challis National Forest, providing insights to guide fire management strategies. Building on this insightful case study, future endeavors aim to extend this methodology to encompass both USA and Canadian regions, with the goal of understanding variations in fire clustering patterns across different geographical areas. Furthermore, considering the significance of climate shifts, there is interest in investigating how altered climate dynamics influence the intricate connections between various parameters and wildfire behavior.



\section{Data availability}
The entire data set used for the analysis shown in this study is available for public use in \cite{cwfis} website.
\section*{Acknowledgments}
I would like to thank Dr. Sina Kazemian for his useful comments. 


\bibliographystyle{tfcad}
\bibliography{interactcadsample}

\begin{thebibliography}{31}
\newcommand{\enquote}[1]{``#1''}
\providecommand{\natexlab}[1]{#1}
\providecommand{\url}[1]{\normalfont{#1}}
\providecommand{\urlprefix}{}

\bibitem[snc(2023)]{sncf}
 2023. ``{SNCF}.'' \url{https://www.fs.usda.gov/scnf}. [Online; accessed
  21-July-2023].

\bibitem[Bedia et~al.(2014)]{bedia2014forest}
Bedia, Joaqu{\'\i}n, Sixto Herrera, Andrea Camia, Jose~Manuel Moreno, and
  Jose~Manuel Guti{\'e}rrez. 2014. ``Forest fire danger projections in the
  Mediterranean using ENSEMBLES regional climate change scenarios.''
  \emph{Climatic Change} 122: 185--199.

\bibitem[Breiman(2001)]{breiman2001random}
Breiman, L. 2001. ``Random forests.'' \emph{Machine learning} 45.

\bibitem[Carvalho et~al.(2008)]{carvalho2008fire}
Carvalho, Anabela, Mike~D Flannigan, K~Logan, Ana~Isabel Miranda, and Carlos
  Borrego. 2008. ``Fire activity in Portugal and its relationship to weather
  and the Canadian Fire Weather Index System.'' \emph{International Journal of
  Wildland Fire} 17 (3): 328--338.

\bibitem[Certini(2005)]{certini2005effects}
Certini, Giacomo. 2005. ``Effects of fire on properties of forest soils: a
  review.'' \emph{Oecologia} 143: 1--10.

\bibitem[Chuvieco et~al.(2016)]{chuvieco2016new}
Chuvieco, Emilio, Chao Yue, Angelika Heil, Florent Mouillot, Itziar
  Alonso-Canas, Marc Padilla, Jose~Miguel Pereira, Duarte Oom, and Kevin
  Tansey. 2016. ``A new global burned area product for climate assessment of
  fire impacts.'' \emph{Global Ecology and Biogeography} 25 (5): 619--629.

\bibitem[CWFIS(2023)]{cwfis}
CWFIS, The. 2023. ``{Canadian Wildland Fire Information System}.''
  \url{https://cwfis.cfs.nrcan.gc.ca/downloads/hotspots/}. [Online; accessed
  21-July-2023].

\bibitem[Dennison et~al.(2014)]{dennison2014large}
Dennison, Philip~E, Simon~C Brewer, James~D Arnold, and Max~A Moritz. 2014.
  ``Large wildfire trends in the western United States, 1984--2011.''
  \emph{Geophysical Research Letters} 41 (8): 2928--2933.

\bibitem[Ester et~al.(1996)]{ester1996density}
Ester, Martin, Hans-Peter Kriegel, J{\"o}rg Sander, Xiaowei Xu, et~al. 1996.
  ``A density-based algorithm for discovering clusters in large spatial
  databases with noise.'' In \emph{kdd}, Vol.~96, 226--231.

\bibitem[Farhani, Sica, and Daley(2021)]{farhani2021classification}
Farhani, Ghazal, Robert~J Sica, and Mark~Joseph Daley. 2021. ``Classification
  of lidar measurements using supervised and unsupervised machine learning
  methods.'' \emph{Atmospheric Measurement Techniques} 14 (1): 391--402.

\bibitem[Finney et~al.(1994)]{finney1994farsite}
Finney, Mark~A, et~al. 1994. ``FARSITE: a fire area simulator for fire
  managers.'' In \emph{the Proceedings of The Biswell Symposium, Walnut Creek,
  California}, 7. Citeseer.

\bibitem[Flannigan and Harrington(1988)]{flannigan1988study}
Flannigan, MD, and JB~Harrington. 1988. ``A study of the relation of
  meteorological variables to monthly provincial area burned by wildfire in
  Canada (1953--80).'' \emph{Journal of Applied Meteorology and Climatology} 27
  (4): 441--452.

\bibitem[Flannigan et~al.(2013)]{flannigan2013global}
Flannigan, Mike, Alan~S Cantin, William~J De~Groot, Mike Wotton, Alison
  Newbery, and Lynn~M Gowman. 2013. ``Global wildland fire season severity in
  the 21st century.'' \emph{Forest Ecology and Management} 294: 54--61.

\bibitem[Fraser, Li, and Cihlar(2000)]{fraser2000hotspot}
Fraser, RH, Z~Li, and J~Cihlar. 2000. ``Hotspot and NDVI differencing synergy
  (HANDS): A new technique for burned area mapping over boreal forest.''
  \emph{Remote sensing of environment} 74 (3): 362--376.

\bibitem[Freyberg et~al.(2022)]{freyberg2022idaho}
Freyberg, Ford, Brenner Burkholder, Jessica Hiatt, and Carson Schuetze. 2022.
  ``Idaho Wildfires: Assessing Drought and Fire Conditions, Trends, and
  Susceptibility to Inform State Mitigation Efforts and Bolster Monitoring
  Protocol in North-Central Idaho.''  .

\bibitem[Giannaros, Kotroni, and Lagouvardos(2021)]{giannaros2021climatology}
Giannaros, Theodore~M, Vassiliki Kotroni, and Konstantinos Lagouvardos. 2021.
  ``Climatology and trend analysis (1987--2016) of fire weather in the
  Euro-Mediterranean.'' \emph{International Journal of Climatology} 41:
  E491--E508.

\bibitem[Green et~al.(1995)]{green1995fire}
Green, Kass, Mark Finney, Jeff Campbell, David Weinstein, and Vaughan Landrum.
  1995. ``Fire! using GIS to predict fire behavior.'' \emph{Journal of
  Forestry} 93 (5): 21--25.

\bibitem[Halladin-DKbrowska, Kania, and Kope{\'c}(2019)]{halladin2019t}
Halladin-DKbrowska, Anna, Adam Kania, and Dominik Kope{\'c}. 2019. ``The t-SNE
  algorithm as a tool to improve the quality of reference data used in accurate
  mapping of heterogeneous non-forest vegetation.'' \emph{Remote Sensing} 12
  (1): 39.

\bibitem[Hastie, Tibshirani, and Friedman(2009)]{unsupervised}
Hastie, T., R.~Tibshirani, and J.~Friedman. 2009. ``Unsupervised learning.'' In
  \emph{The elements of statistical learning}, 485--585.

\bibitem[Holden et~al.(2018)]{holden2018decreasing}
Holden, Zachary~A, Alan Swanson, Charles~H Luce, W~Matt Jolly, Marco Maneta,
  Jared~W Oyler, Dyer~A Warren, Russell Parsons, and David Affleck. 2018.
  ``Decreasing fire season precipitation increased recent western US forest
  wildfire activity.'' \emph{Proceedings of the National Academy of Sciences}
  115 (36): E8349--E8357.

\bibitem[Kasischke and Turetsky(2006)]{kasischke2006recent}
Kasischke, Eric~S, and Merritt~R Turetsky. 2006. ``Recent changes in the fire
  regime across the North American boreal region—Spatial and temporal
  patterns of burning across Canada and Alaska.'' \emph{Geophysical research
  letters} 33 (9).

\bibitem[Koutsias et~al.(2012)]{koutsias2012relationships}
Koutsias, Nikos, Gavriil Xanthopoulos, Dimitra Founda, Fotios Xystrakis, Foula
  Nioti, Magdalini Pleniou, Giorgos Mallinis, and Margarita Arianoutsou. 2012.
  ``On the relationships between forest fires and weather conditions in Greece
  from long-term national observations (1894--2010).'' \emph{International
  Journal of Wildland Fire} 22 (4): 493--507.

\bibitem[Kreye et~al.(2014)]{kreye2014fire}
Kreye, Jesse~K, Nolan~W Brewer, Penelope Morgan, J~Morgan Varner, Alistair~MS
  Smith, Chad~M Hoffman, and Roger~D Ottmar. 2014. ``Fire behavior in
  masticated fuels: A review.'' \emph{Forest Ecology and Management} 314:
  193--207.

\bibitem[Maaten and Hinton(2008)]{maaten2008visualizing}
Maaten, L., and G.~Hinton. 2008. ``Visualizing data using t-SNE.''
  \emph{Journal of machine learning research} 9: 2579--2605.

\bibitem[McWethy et~al.(2019)]{mcwethy2019rethinking}
McWethy, David~B, Tania Schoennagel, Philip~E Higuera, Meg Krawchuk, Brian~J
  Harvey, Elizabeth~C Metcalf, Courtney Schultz, et~al. 2019. ``Rethinking
  resilience to wildfire.'' \emph{Nature Sustainability} 2 (9): 797--804.

\bibitem[Robichaud(2000)]{robichaud2000evaluating}
Robichaud, Peter~R. 2000. \emph{Evaluating the effectiveness of postfire
  rehabilitation treatments}. US Department of Agriculture, Forest Service,
  Rocky Mountain Research Station.

\bibitem[Shvidenko et~al.(2011)]{shvidenko2011impact}
Shvidenko, AZ, DG~Shchepashchenko, EA~Vaganov, AI~Sukhinin, Sh~Sh Maksyutov,
  I~McCallum, and IP~Lakyda. 2011. ``Impact of wildfire in Russia between
  1998--2010 on ecosystems and the global carbon budget.'' In \emph{Doklady
  Earth Sciences}, Vol. 441, 1678--1682. Springer.

\bibitem[Song et~al.(2019)]{song2019improved}
Song, Weijing, Lizhe Wang, Peng Liu, and Kim-Kwang~Raymond Choo. 2019.
  ``Improved t-SNE based manifold dimensional reduction for remote sensing data
  processing.'' \emph{Multimedia Tools and Applications} 78: 4311--4326.

\bibitem[Stocks et~al.(1989)]{stocks1989canadian}
Stocks, Brian~J, BD~Lawson, ME~Alexander, CE~Van Wagner, RS~McAlpine,
  TJ~Lynham, and DE~Dube. 1989. ``The Canadian forest fire danger rating
  system: an overview.'' \emph{The Forestry Chronicle} 65 (6): 450--457.

\bibitem[Van~Wagner et~al.(1987)]{van1987development}
Van~Wagner, CE, et~al. 1987. \emph{Development and structure of the Canadian
  forest fire weather index system}. Vol.~35.

\bibitem[Wotton(2009)]{wotton2009interpreting}
Wotton, B~Mike. 2009. ``Interpreting and using outputs from the Canadian Forest
  Fire Danger Rating System in research applications.'' \emph{Environmental and
  ecological statistics} 16: 107--131.

\end{thebibliography}


\end{document}